\documentclass{icrc2009}

\usepackage{graphicx}   
\usepackage[caption=false]{caption}    
\usepackage[font=footnotesize]{subfig} 
\usepackage{fixltx2e}
\usepackage{url}

\newcommand{\shorttitle}[1]%
{\markboth{Proceedings of the 31\MakeLowercase{$^{st}$} ICRC, {\L}\'{o}d\'{z} 2009}{#1} }

\begin{document}
\title{The Central Control of the MAGIC telescopes.}

\author{\IEEEauthorblockN{R.~Zanin\IEEEauthorrefmark{1} and
 			                      J.~Cortina\IEEEauthorrefmark{1} for the MAGIC Collaboration}\\

\IEEEauthorblockA{\IEEEauthorrefmark{1}IFAE, Edifici Cn., Campus UAB, E-08193 Bellaterra, Spain}
}

\shorttitle{The Central Control of the MAGIC telescopes.}
\maketitle

\begin{abstract}
The second MAGIC telescope, a clone of the first 17 m diameter MAGIC telescope, 
has entered the final commissioning phase and will soon start to take data, 
preferentially in the so-called stereo-mode. The control system for both telescopes 
is assigned to a number of autonomous functional units called subsystems. 
The control hardware and software components of the second telescope 
subsystems have been modified with respect to their counterparts of the first telescope. 
A new Central Control (CC) program has been developed to communicate 
with all the subsystems of both telescopes and to coordinate their functionality 
thus easing the stereo data taking procedure. We describe the whole control 
system in detail: all the subsystems and their communication with the Central 
Control, the CC graphic user interface that grants operators 
the full control over the two telescopes, and the automatic checking 
procedures, which guarantee the safety and 'health' of the apparatus.
\end{abstract}

\begin{IEEEkeywords}
control system ---
MAGIC telescopes ---
stereoscopic system  
\end{IEEEkeywords}

 \section{Introduction}
The MAGIC telescopes consist in two 17~m diameter 
Imaging Atmospheric Cherenkov Telescopes (IACTs) located in the
Canary Island of La Palma ($28^{\circ}$ 45'N, $17^{\circ}$ 54'W, 2225~m a.s.l.).   
MAGIC-I is in operation since 2004, while MAGIC-II is still under 
commissioning, even though it is already taking data.\\
The MAGIC telescopes, designed for the detection of Very High Energy 
(VHE) $\gamma$-rays, have the largest light collection area, and consequently 
the lowest trigger energy threshold among the current generation 
of IACTs. The stereoscopic observation mode will lead to an improvement
in sensitivity by a factor 2 raising new challenges on the pulsar physics,
on the detection of more distant AGNs, and on physical 
studies on weaker galactic sources. \\
Each telescope is composed of different autonomous components controlled 
by their control programs. These functional units, described in section \ref{control}, 
communicate with the Central Control (CC) program which coordinates the 
actions of each telescope in order to automatize
the stereo data taking procedure as much as possible (see section \ref{CC}). Nevertheless,
the CC graphic user interface confers the full control over the telescopes 
on the operators. For example, they can configure the various components 
of the telescopes according to the observation necessities. They can also stop 
every kind of automatic procedure in case of problems during the data taking. \\
The role of the CC program is fundamental: it provides the user with a 
simple interface to operate the telescopes in an orderly manner. On the other hand, 
it defines all the standard observational procedures and the needed checks which
guarantee the safety and "health" of the apparatus, as presented in section \ref{safety}.\\

\section{The Control System} \label{control}
The control system of the MAGIC telescopes is split up into 
autonomous functional units, called subsystems. They control 
the different components the telescopes are composed of: the telescope
drive system, the readout, the camera, the Active Mirror Control (AMC),
the data acquisition system and the calibration. The MAGIC-II subsystems
were upgraded with respect to the first telescope ones from either a 
hardware or a software point of view.  Some existing
problems were solved and their performance improved. 
Moreover, there are the so-called "common subsystems" which 
collect information useful for the data taking, like the weather 
conditions, the GRB alerts sent by different satellites, and the 
absolute time given with high precision. All the subsystems, 
classified according to their telescope belonging, are
described below.\\
\vspace{0.2cm}

The MAGIC-I subsystems are:
\begin{itemize}
\item{\bf Telescope Drive System 1:}
it steers two ALT/AZ motors and monitors the telescope position 
using two shaft encoders and two rotary encoders over CANBus \cite{drive}.
\item{\bf Active Mirror Control 1:}
two motors behind each $\mathrm{1m^{2}}$ mirror panel 
(4$\times$0.5$\mathrm{m^{2}}$ mirrors) allow to 
counteract any deformation of the mirror support dish during 
repositioning and tracking. Each panel is equipped with a guidance 
laser in order to monitor the panel orientation. Moreover, the AMC 
includes four LEDs on the camera lid and a CCD-camera mounted 
on the reflector frame close to its center. All these component are 
connected to the control pc over RS-485 bus \cite{amc}. 
\item{\bf Camera and Calibration 1:} the MAGIC-I camera comprises 576
Photo MultiPliers (PMTs) of two different sizes. The inner part of the camera is covered by 
396 $0.1^{\circ}$ Field Of View (FOV) PMTs. The outer part is composed by
180 $0.2^{\circ}$ FOV PMTs \cite{camera1}.
The communication with the monitoring and configuring system of the pixels, 
and the calibration \cite{calibration} is done via CANBus lines. The camera includes also a 
cooling system, two lids which cover the camera
during day light and low voltages power. These three electronic components are
autonomously controlled using several PLCs (Programmable Logic Controller) and 
the access to those PLCs is done using a protocol over RS-485 called Modbus.
\item{\bf Trigger system:} The analog PMT signals are transmitted via optical fibers and divided
by optical splitters at the counting house. Half of the signal is sent to the receiver boards which 
shape the signals through discriminator thresholds and send them to the level 1 trigger 
via Low Voltage Differential Signal (LVDS) system. The level 1 trigger is based on  fast N-fold 
(N=2-5) next neighbor (NN) logic \cite{trigger}. 
The triggered signals are then sent via LVDS to a VME system, connected through 
VME bus to the control pc, which allows to monitor and set the trigger parameters. 
\item{\bf Mux DAQ:} After having split the PMT signals, the second part of them is sent 
to the fiber-optic multiplexing readout system. The latter uses a 2GSample/s Flash 
Analog to Digital Converter (FADC) to digitize 16 channel consecutively. Within group 
of 16 channels, each analog signal is delayed of 40 ns with respect to the previous one by 
using optical fibers. These signals are then converted to electronic 
ones by using PIN diodes. Fast high bandwidth MOSFET-switches 
open for 40 ns channel by channel. Finally the signals are summed in two active summation 
stages and digitized by the FADC \cite{MUX}.
\item{\bf Sum Trigger:} it is a second trigger system based on a new analog trigger concept 
which allows to reduce significantly the energy threshold. This trigger divides the camera in 
24 semi-overlapping patches consisting in 18 pixels each. 
The patch analog signals are clipped at a certain amplitude to prevent 
accidental triggers from PMT-afterpulses, and then summed up. 
The analog sum signals are compared to programmable analog thresholds 
in semi-customized boards. The host boards communicate with a VME backplane for the readout
of the patch rates and the threshold settings via a USB-VME bridge \cite{ST}.
\end{itemize}
\begin{figure*}[!th]
\begin{center}
   \includegraphics[width=5in]{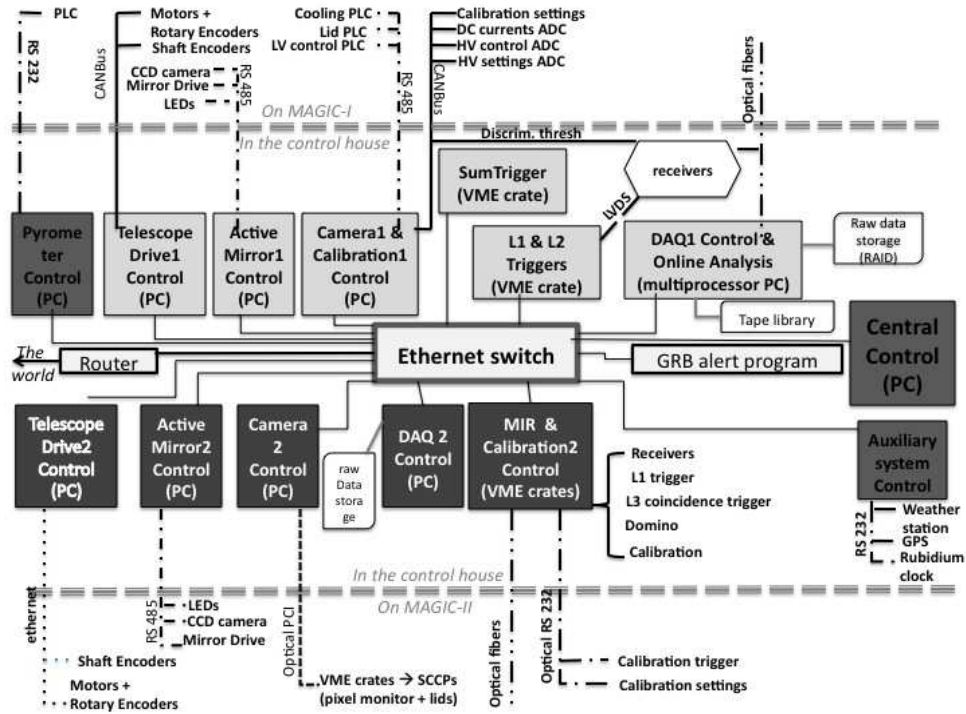}
   \caption{Outline of the MAGIC control system.}
   \label{scheme}
   \end{center}
 \end{figure*}
The MAGIC-II subsystems are:
\begin{itemize}
\item{\bf Telescope Drive System 2:}
it steers two ALT/AZ motors and monitors the telescope position 
using a shaft encoder and two rotary encoders which communicate with a
PLC via Profibus DP. The PLC is connect to the control program via ethernet 
\cite{drive}.
\item{\bf Active Mirror Control 2:} it works like in MAGIC-I, even though, 
instead of mirror panels, there are single 1$\mathrm{m^{2}}$ mirrors. This subsystem
was improved a lot at the software level \cite{amc}.
\item{\bf Camera 2:} the camera of MAGIC-II is uniformly equipped 
with 1039 $0.1^{\circ}$ diameter PMTs \cite{camera2}.
The pixels are grouped into clusters of seven pixels each. Each cluster 
is controlled by a Slow Control Cluster Processor (SCCP) which monitors
and sets all the PMT-related parameters. Also the camera lids are steered by SCCPs. 
These SCCPs are connected via LAN cables to a VME crate inside the 
camera box, which communicates to the camera control PC in the counting 
house via an optical PCI to VME link \cite{caco2}.
\item{\bf Readout system, calibration and trigger:} from a hardware point of view 
the readout system is completely
different from the MAGIC-I one. The PMT analog signals, transmitted by optical fibers
to the counting house, are converted to electronic ones and split 
into two branches inside the receiver boards: one 
provides the input for the level one trigger (VME crate), the second one is sent 
 to digitizing units. The triggered Cherenkov pulses are sampled at
 2GSample/s by a Domino Ring Chip and temporarily stored in 
 1024 capacitors. The Domino chip is installed in the PULSAR 
 boards which are VME interface boards \cite{readout}. A slow control 
 program, called Magic Integrated Readout (MIR), is a multi-thread
 C-program which communicates with
 all the readout hardware components through the VME bus. The MIR
 steers also the calibration box which is connected to a VME crate in the
 counting house using a RS232 protocol over optical cables.   
\item{\bf DAQ 2:} it is a multithread C++ program, based on MAGIC-I
Mux Readout, which reads the data samples from the readout hardware,
performs an online analysis of the data and finally, stores the data in 
dedicated disk \cite{readout}. The achievable acquisition rate is $\sim$1kHz.
\end{itemize}
The "common" subsystems are:
\begin{itemize}
\item{\bf Pyrometer:}  Germanium lenses focus the
radiation on thermoelectric sensor which measures the radiation
temperature. Its integral flux defines the sky temperature. Another
small sensor measures the air temperature and the humidity. This
device, installed on the left side of the MAGIC-I mirror surface, is 
controlled by a PLC which communicates with the control program
using a protocol RS-232. 
\item{\bf GRB alert program:} it is a daemon which monitors 
the Gamma-ray burst Coordinates Network (GCN) for any 
alert on GRBs. These alerts are promptly noticed to the 
Central Control providing the "trigger" for the repositioning 
of the telescope.
\item{\bf Auxiliary PC:} it collects information sent over a serial RS232 connection 
by a GPS receiver, a Rubidium atomic clock and the weather station. 
The two clocks provide the time stamp for the DAQs. The difference between 
both devices is monitored through a NIM module and reported 
to the auxiliary PC. The values are then written into a file. \\
The weather station is a microprocessor Weather Station MWS 5MV high quality steel 
with datalogger and forced ventilation by fan made by Reinhardt GmbH. 
The information collected by this device is read out over a serial connection by
a program which writes the weather values onto a file.
\item{\bf Level 3 coincidence trigger:} it is the coincidence trigger of the two telescopes 
made of VME boards which are controlled by the MIR program. 
\end{itemize}
\hspace{0.2cm}

\noindent {All the subsystems can be monitored and configured through 
their control programs which can be run in stand-alone mode for testing 
purposes, or remotely through the Central Control (CC) program. 
Each control program is installed in its correspondent PC in the 
counting house which is connected to the ethernet switch of the 
internal network. The scheme of the MAGIC control system is shown 
in figure \ref{scheme}.}

\hspace{5cm}

\section{The Central Control} \label{CC}
The central control program is implemented in a Linux PC and written in LabView 
8.5. It provides the user an easy graphical interface to operate the two 
MAGIC telescopes. \\
The central control is the interface between all the subsystems which cannot interact 
with each other. It communicates directly with each of them via socket connection, expect 
in the case of  the weather station and the clocks  for which a file sharing is used.
At startup, it starts polling the other subsystems for availability. As soon as they are up, 
the communication is established and this fact is displayed to the user.  \\
Every subsystem sends an ASCII report to CC over ethernet using the TCP/IP protocol 
every second describing its state, time stamp, and all relevant operation parameters.
This state variable is used to define the action of the subsystem. If 
the central control does not receive any report from a subsystem for more than 10 s, 
it defines the correspondent subsystem state as "unavailable". 
By permanently updating the global subsystems state vectors, CC  
keeps track of the state of the two telescopes and perform the needed actions 
accordingly. 
\begin{figure*}[!]
\begin{center}
    \includegraphics[width=4in]{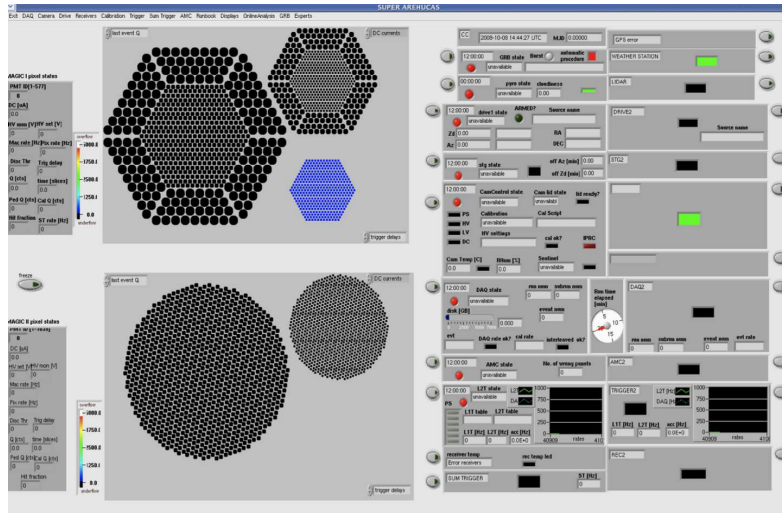}
    \caption{GUI of the central control program.}
   \label{GUI}
   \end{center}
 \end{figure*}
The received information, which is useful to determine the performance of the 
telescopes during data taking, is displayed in the panel on the right 
side of the Graphical User Interface (GUI), as shown in figure \ref{GUI}. 
The top part of this panel includes all the "common subsystems", while below 
all the information of MAGIC-I and MAGIC-II subsystems respectively are shown. 
In particular, every subsystem has its own window inside the panel. 
On the other hand, the two camera displays, on the left side of the GUI, can show 
the information related to each pixel (voltages, currents, discriminator thresholds, 
pulse arrival time, pulse charge, etc).
Moreover, the reports received by the central control during a data run are 
written in a specific file and later on used as additional information for the offline analysis.\\
The central control sends a report every 10s to all the subsystems which, in this way, 
can know when CC is available. Since the subsystems cannot communicate with
each other, the CC report is the only way they have to be aware of the state and some
useful parameters from a different subsystem. \\
Commands from CC to the other subsystems are executed by them as if the user 
had given them directly at the consoles of the subsystems. In order to avoid accidental 
interference, the subsystems, when controlled by CC, are locked from direct
interaction at their console. \\
The central control program contains routines which perform different actions,
needed for the data taking or for testing purposes, by sending a sequence of commands
to the involved subsystems. This automatization makes the operation of the telescopes
much easier and prevents possible human errors. Nevertheless, CC grants the 
user full control of the telescopes, by allowing him to configure the system, to
stop any automatic procedure in case of problems and to skip it in case of particular
circumstances by sending single commands to each subsystem.

\subsection{Automatic Checking Procedures} \label{safety}
The central control performs decisional loops before sending any command in order to
check that the involved subsystem is in the correct configuration. These checks prevent 
possible mistakes which can deteriorate the data quality and avoid that the command 
will be unsuccessful. \\ 
Moreover the parameters delivered by the subsystems are compared to safety limits or standard 
values and warning or safety reactions result when the parameters seem to show problematic states. 
For example, if the external humidity, measured by the weather station, exceeds a certain value, the
electronic inside the camera can be damaged. Therefore, the lids, which protect the camera, will
be automatically closed. If the sun is still present, the system does not allow the operator
to rump up the high voltage in order not to ruin the PMTs. If the wind speed is too high, this can provoke
strong damages to the telescope structure and to camera lids if they are opened. In this case CC closes
the lids and park the telescope.\\ 

\section{Acknowledgement} \label{conclusion}
We would like to thank the Instituto de Astrofisica Canario (IAC) for the excellent
working conditions at the Observatorio del roque de los Muchachos in La Palma.
The support of the German BMBF and MPG, the Italian INFN, and Spanish MCINN
is gratefully acknowledged. This work is also supported by the ETH Research Grant
TH 34/043, and by the YIP of the Helmholtz Gemeinschaft. The authors would also like
to thanks all the MAGIC colleagues who made this work possible.


\begin{thebibliography}{99}
 
%
%
%
%
\bibitem{drive} Bretz, T. et al. The drive system of the Major Atmospheric Gamma-ray Imaging Atmospheric Telescope. Astropart.Phys.31:92-101,2009. 
\bibitem{amc} Biland, A. et al. The Active Mirror Control of the MAGIC Telescopes. 30th ICRC Proceedings. 
\bibitem{camera1} Cortina, J. et al. Camera Control and Central Control of the MAGIC Telescope. 28th ICRC Proceedings.
\bibitem{calibration} Gaug, M. et al. An Absolute Light Flux Calibration for the MAGIC Tele-
scope. 28th ICRC Proceedings.
\bibitem{trigger} Paoletti, R. The Trigger System of the MAGIC Telescope. IEEE transactions on nuclear science, vol. 54, no. 2, April 2007 
\bibitem{MUX} Goebel, F. et al. Upgrade of the MAGIC Telescope with a Multiplexed Fiber-Optic 2 GSamples/s FADC Data Acquisition system. 28th ICRC Proceedings
\bibitem{ST} Rissi, M. et al. A New Trigger Provides a Lower Energy Threshold for the MAGIC Cherenkov Telescope. Nuclear Science Symposium '08 proceedings.
\bibitem{camera2} Borla-Tridon, D. et al. Performances of the Camera of the MAGIC-II Telescope. These proceedings.
\bibitem{caco2} Steinke, B. and Jogler, T. MAGIC-II Camera Slow Control Software. These Proceedings.
\bibitem{readout} Tescaro, D. et al. The readout system of the MAGIC-II Cherenkov Telescope. These proceedings.
\end{thebibliography}
\end{document}